\documentclass[aps,pra,superscriptaddress,floatfix,showpacs,10pt, twocolumn]{revtex4}
\usepackage[dvips]{graphicx}
\usepackage{amsmath}

\begin{document}

\title{Entanglement evolution of a two-qubit system with decay beyond rotating-wave approximation}
\author{Qing Yang}
\email{qingyang@ahu.edu.cn}\affiliation{School of Physics {\&}
Material Science, Anhui University, Hefei, 230039}
\author{Ming Yang}
\affiliation{School of Physics {\&} Material Science, Anhui
University, Hefei, 230039}
\author{Da-Chuang Li}
\affiliation{School of Physics {\&} Material Science, Anhui
University, Hefei, 230039}
\author{Zhuo-Liang Cao}
\email{zlcao@ahu.edu.cn}\affiliation{ Department of Physics, Hefei
Teachers College,230061}

\begin{abstract}
Two noninteracting atoms, initially entangled in Bell states, are
coupled to a one-mode cavity. Based on the reduced non-perturbative
quantum master equation, the entanglement evolution of the two atoms
with decay is investigated beyond rotating-wave approximation. It is
shown that the counter-rotating wave terms have great influence on
the disentanglement behavior. The phenomenon of entanglement sudden
death and entanglement sudden birth will occur.
\end{abstract}

\maketitle
\section{introduction}
Entanglement plays a central role in the field of quantum
information science where it is considered as a valuable resource
for some non-classical tasks, such as: quantum
computation\cite{comp1,comp2,comp3}, quantum teleportation
\cite{tele}, superdense coding\cite{super}, and quantum cryptography
\cite{cryp1,cryp2}. Applications of interest have triggered research
on the dynamical behavior of entanglement in order to control
quantum disentanglement\cite{8,9,10}.

Recently, the dynamics of entanglement in bi-partite systems has
been under extensive research \cite{11,12,13,14,15,16,17,18,19,20}
since the work of Yu and Eberly\cite{ESD}, in which the entanglement
may terminate abruptly in a finite time while coherence is lost
asymptotically. This effect is termed entanglement sudden
death(ESD). Subsequently, C.E.Lo'pez \emph{etc.} put forward a new
term "entanglement sudden birth"(ESB)\cite{birth} and try to present
an explanatory study of multipartite entanglement evolution.

In the previous studies, the rotating-wave approximation(RWA), which
neglecting counter rotating terms corresponding to the emission and
absorption of virtual photon without energy conservation, is widely
used. Generally, the coupling ratio of the atom-field is of the
order $10^{-7}\sim10^{-6}$ in atom-field cavity systems and the RWA
is justified. Recently, D.Meiser \emph{etc.} have investigated the
cavity systems with superstrong coupling\cite{23}. It can be seen
that the effect of counter-rotating terms should be considered in
the strong coupling regime, such as solid state systems.

Different from the previous works, we study the disentanglement
between a pair of qubits with decay which interacting with a cavity
beyond the RWA. Considering the real processing, the atomic decay is
unavoidable, which can be caused by two different physical
mechanisms. The first is the transition without light radiation. In
this process the energy is emitted by thermal energy or the other
forms. The second is the radiation transition. There will emit
photons with the atomic transition. In our model the second is the
main physical mechanism because the collision probability between
the two atoms is very small. Thus the discussion upon the effect of
atomic decay caused by spontaneous emission to the entanglement
evolution is necessary.

The paper is organized as follows. In section II, the two-qubit
model with decay is presented and the reduced non-perturbative
quantum master equation of atoms is derived. In section III, the
entanglement evolution of two initially entangled atoms is
investigated by using Wootters' concurrence\cite{concurrence}. In
the last section, the conclusions are given.

\section{Model}

Consider two noninteracting atoms A and B, which are interacting
with a single-mode cavity resonantly. For simplicity, we assume that
the two atoms have the same parameters. It can be described by the
following Hamiltonian:
\begin{equation}
\label{1} H=H_{a}+H_{f}+H_{af},
\end{equation}
where

\begin{equation}
\label{2} H_{a}=\frac{\omega_{0}}{2}\Sigma^{2}_{i=1}\sigma_{z_{i}},
\end{equation}
\begin{equation}
\label{3} H_{f}=\omega a^{+}a,
\end{equation}
\begin{equation}
\label{4} H_{af}=\lambda\sigma(a^{+}+a),
\end{equation}
where  $\sigma=\Sigma^{2}_{i=1}(\sigma^{+}_{i}+\sigma^{-}_{i})$,
$\sigma_{z_{i}}=|e_{i}\rangle\langle e_{i}|-|g_{i}\rangle\langle
g_{i}|$ and $\sigma_{i}^{-}=|g_{i}\rangle \langle e_{i}|$ are the
atomic operators with $|e_{i}\rangle$ and $|g_{i}\rangle$ being the
excited and ground states of the $i$th atom; $\omega_{0}$ is the
atomic transition frequency between the ground state and the excited
state; $\lambda$ is the coupling constant between atom and cavity;
$a$ and $a^{\dag}$ denote the annihilation and creation operators of
the cavity field mode corresponding frequency $\omega$.

Taking the atomic spontaneous emission into consideration, the
reduced non-perturbative quantum master equation of atoms could be
obtained by path integrals\cite{25},
\begin{eqnarray}
\label{5} \frac{\partial}{\partial t}\rho_{a}=&-&i\mathcal
{L}_{a}\rho_{a}-\int^{t}_{0}ds\langle\mathcal {L}_{af}e^{-i\mathcal
{L}_{0}(t-s)} \mathcal {L}_{af}e^{-i\mathcal
{L}_{0}(s-t)}\rangle_{f}\rho_{a} \nonumber\\
&+&\frac{\gamma}{2}\Sigma^{2}_{i=1}(2\sigma^{-}_{i}\rho_{a}\sigma^{+}_{i}
-\sigma^{+}_{i}\sigma^{-}_{i}\rho_{a}-\rho_{a}\sigma^{+}_{i}\sigma^{-}_{i}),
\end{eqnarray}
where $\mathcal {L}_{0}$, $\mathcal {L}_{a}$ and $\mathcal {L}_{af}$
are Liouvillian operators defined as
\begin{eqnarray}
\label{6} \mathcal {L}_{0}\rho=[H_{a}+H_{f},\rho],
\nonumber\\
\mathcal {L}_{a}\rho=[H_{a},\rho], \nonumber\\\mathcal
{L}_{af}\rho=[H_{af},\rho],
\end{eqnarray}
and $<\cdots>_{f}$ stands for partial trace of cavity mode; $\gamma$
is the atomic decay constant.

Assuming that the cavity field is initially in vacuum state, the
non-perturbative reduced master equation of the atoms could be
derived from Eq.(5),
\begin{eqnarray}
\label{7} \frac{\partial}{\partial
t}\rho_{a}=&-&i\frac{\omega_{0}}{2}[\sigma_{z_{1}}+\sigma_{z_{2}},\rho]
\nonumber\\
&-&\alpha\lambda^{2}\sigma[\sigma^{+}_{1}+\sigma^{+}_{2},\rho]-f\lambda^{2}\sigma[\sigma^{-}_{1}+\sigma^{-}_{2},\rho]
\nonumber\\
&+&\alpha^{*}\lambda^{2}[\sigma^{-}_{1}+\sigma^{-}_{2},\rho]\sigma+f^{*}\lambda^{2}[\sigma^{+}_{1}+\sigma^{+}_{2},\rho]\sigma
\nonumber\\
&+&\frac{\gamma}{2}\Sigma^{2}_{i=1}(2\sigma^{-}_{i}\rho_{a}\sigma^{+}_{i}
-\sigma^{+}_{i}\sigma^{-}_{i}\rho_{a}-\rho_{a}\sigma^{+}_{i}\sigma^{-}_{i}),
\end{eqnarray}
where
\begin{eqnarray}
\label{8} \alpha=\frac{1-exp(-i\Delta t)}{i\Delta},
\nonumber\\
f=\frac{exp(i\delta t)-1}{i\delta}.
\end{eqnarray}
and $\Delta=\omega+\omega_{0}$, $\delta=\omega_{0}-\omega$,
$f^{*}$is conjugate of $f$.

\section{disentanglement}

\subsection{Initial state and the entanglement measurement}
In this section, we will use Wootters' concurrence to quantify the
degree of entanglement\cite{concurrence}. For two qubits, the
concurrence is calculated from the density matrix $\rho$ for qubits
A and B:
\begin{equation}
\label{27}
C(\rho)=max(0,\sqrt{\lambda_{1}}-\sqrt{\lambda_{2}}-\sqrt{\lambda_{3}}-\sqrt{\lambda_{4}}),
\end{equation}
where $\lambda_{i}$ are the eigenvalues of the matrix
\begin{equation}
\label{28}
\varrho=\rho_{AB}(\sigma^{A}_{y}\otimes\sigma^{B}_{y})\rho^{\ast}_{AB}(\sigma^{A}_{y}\otimes\sigma^{B}_{y})
\end{equation}
arranged in decreasing order. Here $\rho^{\ast}_{AB}$ denotes the
complex conjugation of $\rho_{AB}$, and $\sigma_{y}^{A(B)}$ is the
standard Pauli matrix acting in the space of qubit A (or B). The
concurrence varies from $C(\rho)=0$ for an unentangled state to
$C(\rho)=1$ for a maximally entangled state.

Here we restrict our analysis to the initial entangled states
\begin{equation}
\label{27} |\phi\rangle=\alpha |01\rangle+\beta |10 \rangle,
 |\psi\rangle=\alpha |00\rangle+\beta |11 \rangle
\end{equation}
where $\alpha$ is real, $\beta=|\beta|e^{i\delta}$ and
$\alpha^{2}+|\beta|^{2}=1$. For these two entangled states, the
initial atomic density matrix has an "X" structure\cite{26} which is
maintained during the evolution\cite{27}. The reduced density matrix
of the two atoms $\rho_{a}$, in the standard product basis $\mathcal
{B}=\{|1\rangle=|ee\rangle, |2\rangle=|eg\rangle,
|3\rangle=|ge\rangle, |4\rangle=|gg\rangle\}$, could be written as
\begin{equation}
\label{16}\rho_{a}=\left(
  \begin{array}{cccc}
    \rho_{11}&0&0&\rho_{14} \\
    0&\rho_{22}&\rho_{23}&0\\
    0&\rho_{32}&\rho_{33}&0\\
    \rho_{41}&0&0&\rho_{44}\\
  \end{array}
\right),
\end{equation}

The concurrence of $\rho_{a}$ could be obtained
\begin{eqnarray}
\label{8}
C_{\phi}(t)=max\{0,2|\rho_{23}|-2\sqrt{\rho_{11}\rho_{44}}\},
\nonumber\\
C_{\psi}(t)=max\{0,2|\rho_{14}|-2\sqrt{\rho_{22}\rho_{33}}\}.
\end{eqnarray}

\subsection{Numerical results and discussion}
In order to study the counter rotating terms effect on the system,
we investigate the entanglement evolution of two atoms by numerical
calculation. For simplicity, the resonant case $\omega=\omega_{0}$
is considered.

First, we focus on the disentanglement of two qubits with the
initial state of $|\phi\rangle$. The entanglement evolution of two
decayed atoms is shown in Fig.1, Fig.2 and Fig.3 with different
parameters. One could see that the change of $C_{\phi}$ against
$\alpha^{2}$ is symmetrical because of the symmetry of the initial
state $|\phi\rangle$.

When $\omega=\lambda$, Fig.1 shows that the concurrence $C_{\phi}$
changes with the initial value $\alpha^{2}$ and the time $\lambda
t$. It can be seen that the concurrence decreases to zero in a
finite time and  the entanglement undergoes the so-called ESD. The
tendency is similar to that of the above case when
$\omega=3\lambda$, as shown in Fig.2.

As $\omega=10\lambda$, Fig.3 reveals the time evolvement of
concurrence. The tendency is different from that of the above cases.
It is easy to note that the entanglement decays to zero in a finite
time, then revives with small amplitude and disappears permanently
at last. Particularly, when the two atoms are not entangled
initially, namely $\alpha^{2}=0$ or $\alpha^{2}=1$, there is
entanglement sudden birth and its amplitude is larger than that of
the case $0<\alpha^{2}<1$, then the entanglement decreases to zero
eventually due to the atomic decay.

\begin{figure}[tbp]
\includegraphics[scale=0.40,angle=0]{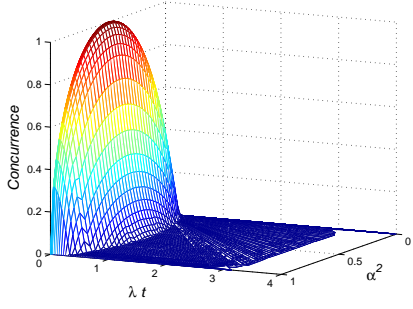}
\caption{The concurrence $C_{\phi}$ as functions of $\alpha^{2}$ and
the time $\lambda t$ with $\omega=\lambda$. } \label{fig1}
\end{figure}
\begin{figure}[tbp]
\includegraphics[scale=0.40,angle=0]{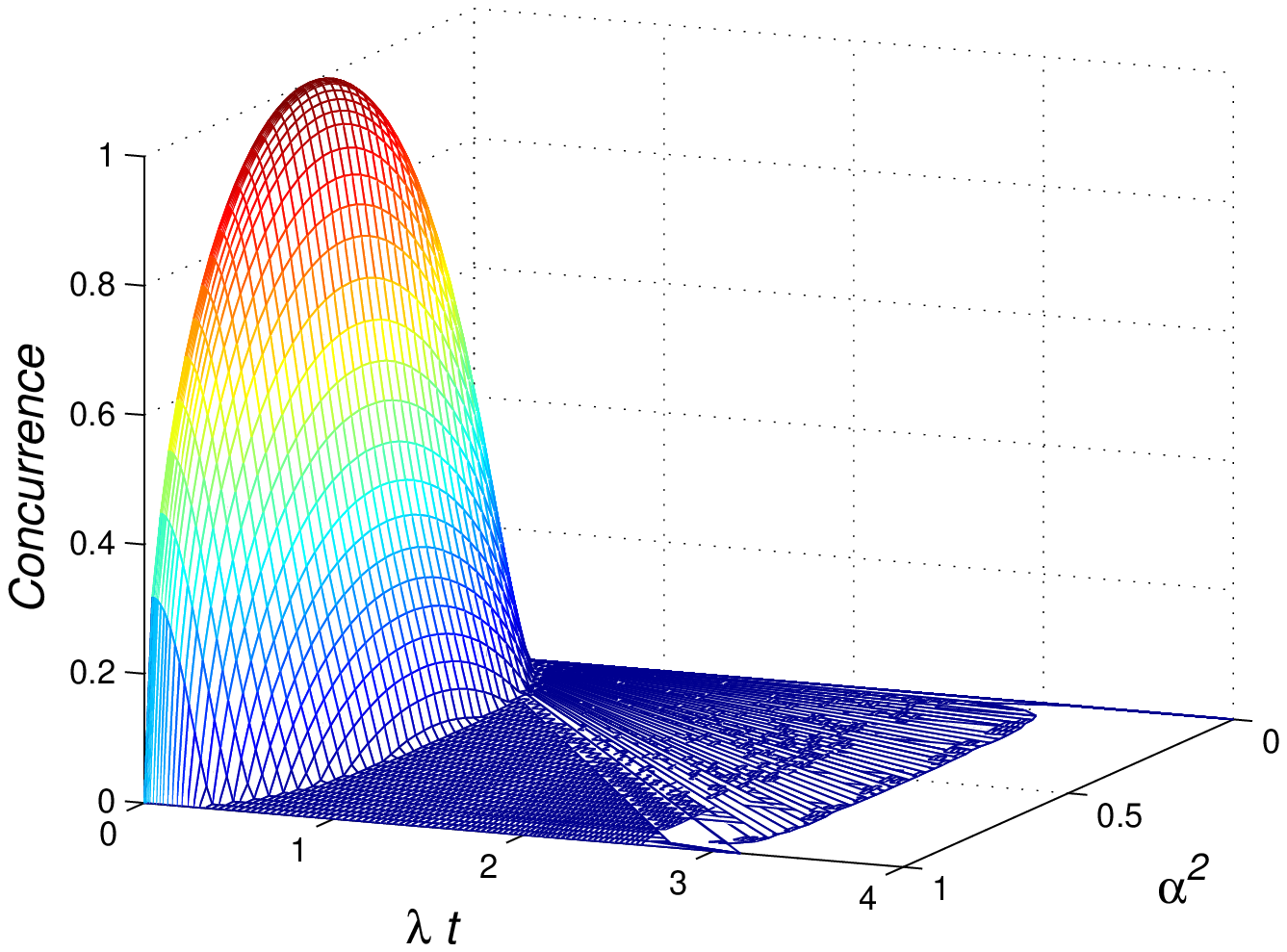}
\caption{The concurrence $C_{\phi}$ as functions of $\alpha^{2}$ and
the time $\lambda_{A}t$ with $\omega=3\lambda$. } \label{fig2}
\end{figure}
\begin{figure}[tbp]
\includegraphics[scale=0.40,angle=0]{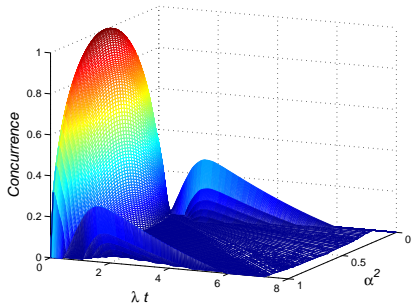}
\caption{The concurrence $C_{\phi}$ as functions of $\alpha^{2}$ and
the time $\lambda_{A}t$ with $\omega=10\lambda$.} \label{fig3}
\end{figure}

Alternatively, we focus on the disentanglement of two qubits with
the initial state $|\psi\rangle$. The entanglement evolvement is
investigated in Fig.4, Fig.5 and Fig.6. The figures are not
symmetrical to $\alpha^{2}$ because the initial state of the two
atoms is asymmetrical.

When $\omega=\lambda$, Fig.4 shows that the concurrence $C_{\psi}$
decays exponentially to zero in almost all cases except that the
value of $\alpha^{2}$ is near $0$. In the case of small $\alpha^{2}$
, there exists small fluctuation before $C_{\psi}$ vanishes
permanently. Particularly, when $\alpha^{2}=0$, there exists small
ESB and ESD. This effect is resulted from the coaction of rotating
wave process and counter-rotating process on the whole system.

By contrast, as $\omega=3\lambda$, Fig.5 reveals that the amplitude
of fluctuation is higher than that of $\omega=\lambda$ in the case
of small $\alpha^{2}$. It is also found that the range existing the
fluctuation is larger than that of the previous case. But there is
no ESB in the case of $\alpha^{2}=0$.

When $\omega=10\lambda$, the time evolution of the concurrence is
plotted in Fig.6. The concurrence decreases monotonically and
exponentially to zero when $\alpha^{2}$ is near $1$. The smaller the
$\alpha^{2}$ is, the more intense the fluctuation is. The
concurrence disappears permanently at last because the atoms decay.

\begin{figure}[tbp]
\includegraphics[scale=0.40,angle=0]{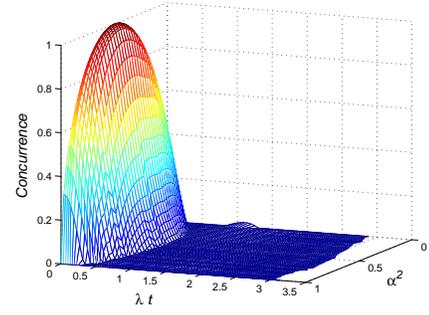}
\caption{The concurrence $C_{\psi}$ as functions of $\alpha^{2}$ and
the time $6^{1/2}\lambda t$ with $\omega=\lambda$. } \label{fig4}
\end{figure}
\begin{figure}[tbp]
\includegraphics[scale=0.45,angle=0]{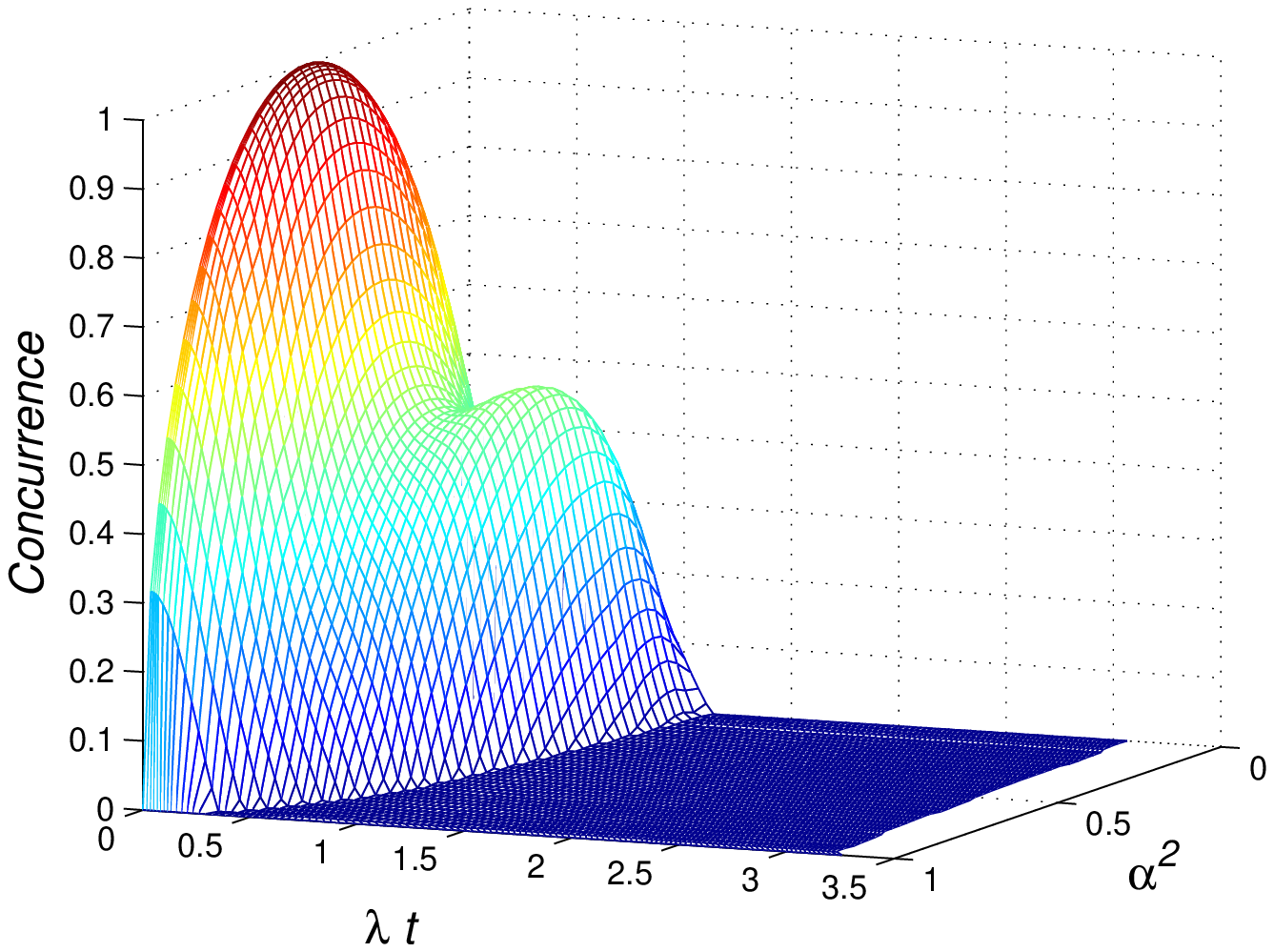}
\caption{The concurrence $C_{\psi}$ as functions of $\alpha^{2}$ and
the time $\lambda t$ for $\omega=3\lambda$.} \label{fig5}
\end{figure}
\begin{figure}[tbp]
\includegraphics[scale=0.45,angle=0]{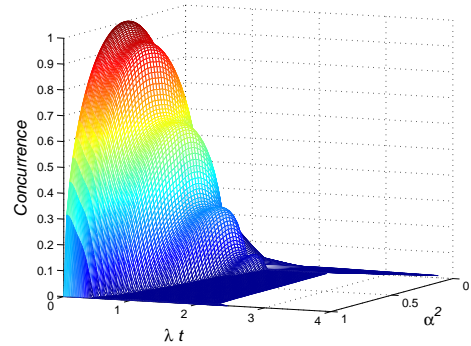}
\caption{The concurrence $C_{\psi}$ as functions of $\alpha^{2}$ and
the time $\lambda t$ for $\omega=10\lambda$.} \label{fig6}
\end{figure}

\section{conclusions}
In summary, we have investigated the disentanglement of two
noninteracting atoms with decay coupling to a one-mode cavity
resonantly beyond RWA. It is shown that the entanglement evolution
is dependent on the ratio of the atom-field coupling divided by the
atomic transition frequency and the phenomena of entanglement
evolution are rich. The physical mechanism behind the phenomena is
the process of emission and absorption of virtual photon and the
atomic decay.

The study of entanglement evolution beyond RWA is a significant
problem because of its importance to the field of strong coupling.
It will help ones to deal with the practical case of solid state
system. The consideration of atomic decay is also of practical
significance.

\begin{acknowledgments}
This work is supported by National Natural Science Foundation of
China (NSFC) under Grant Nos: 60678022 and 10704001, the Specialized
Research Fund for the Doctoral Program of Higher Education under
Grant No. 20060357008, Anhui Provincial Natural Science Foundation
under Grant No. 070412060, the Key Program of the Education
Department of Anhui Province under Grant No. KJ2008A28ZC, and the
Talent Foundation of Anhui University.
\end{acknowledgments}

\end{document}